# Reporte de visita médica en la aplicación de braquiterapia de baja taza con Cs-137 a una paciente con cáncer Cérvix-Cuello Uterino Estadio IIB

Mencomo, Miguel [a] (Lambdamiguel2105@gmail.com)

[a] Estudiante de licenciatura en Física, escuela de Física de la Facultad de Ciencias Naturales Exactas y Tecnología de la Universidad de Panamá, República de Panamá.

**RESUMEN**

*Durante la visita al Instituto Oncológico Nacional (ION), se presenció el proceso de braquiterapia de baja tasa con Cs-137 en un paciente con cáncer cervicouterino estadio 2B. Se observó la meticulosa preparación de la paciente y la planificación de la dosis mediante toma de imágenes. A pesar de las limitaciones logísticas identificadas en el proceso, se enfatizó la importancia de seguir rigurosamente el protocolo de vestimenta, desinfección y manejo de fuentes radiactivas. Se destacó el cuidadoso manejo de las fuentes por parte del técnico y el físico médico. La aplicación de conceptos físicos como los rayos X brindó un valioso aprendizaje, resaltando la relevancia del enfoque interdisciplinario en el tratamiento de braquiterapia. La colaboración entre diversos profesionales de la salud y un ambiente de cooperación fue fundamental para garantizar la seguridad y la efectividad del tratamiento.*

**Palabras claves**:

*Braquiterapia de baja taza, Rayos x, Cáncer, Tándem.*

## 1. Introducción

La braquiterapia, un tipo de radioterapia interna, implica la colocación de fuentes radioactivas de alta energía cerca o incluso dentro del tejido tumoral. Esta técnica se usa comúnmente en el tratamiento del cáncer cervicouterino para administrar dosis altamente focalizadas de radiación directamente al sitio afectado, lo que minimiza el impacto en los tejidos sanos circundantes. La braquiterapia de baja tasa con cesio-137, en particular, se refiere al uso de una fuente radioactiva específica (cesio-137) que se introduce temporalmente en la paciente para emitir radiación de baja tasa y de forma controlada.

La fuente radiactiva de cesio-137 es comúnmente utilizada en aplicaciones médicas, industriales y de investigación debido a su emisión de radiación gamma de alta energía. Antiguamente, se solía medir su actividad en la unidad del Curie, y en el contexto del Sistema Internacional de Unidades (SI), se utiliza el becquerel (Bq) como medida estándar para la actividad radiactiva. Por otro lado, el gray (Gy) se utiliza para cuantificar la absorción de dosis de radiación en un material específico. En el campo de la radioterapia, el cGray (centiGray) es una medida importante para determinar la dosis de radiación absorbida por los tejidos. La relación entre el cGray y el gray es de 1 cGray equivalente a 0.01 Gy.

El término "estadio 2B" en el cáncer cervicouterino describe una etapa en la que el cáncer se ha extendido más allá del cuello uterino hasta los tejidos circundantes, pero aún no ha alcanzado la pared pélvica ni ha afectado la uretra y la vagina. Los estadios del cáncer cervicouterino se definen según la extensión del tumor, y van desde el estadio 0, donde el cáncer está confinado al revestimiento superficial del cuello uterino, hasta los estadios 3 y 4, donde el cáncer se ha diseminado a otras partes del cuerpo.

Aparte del estadio 2B (*fig. 1*), otros estadios incluyen el estadio 1, que se subdivide en estadios 1A y 1B dependiendo del tamaño del tumor, el estadio 2A que implica una invasión ligeramente mayor que el estadio 1, y los estadios 3 y 4, donde el cáncer se ha propagado a la pelvis y otros órganos distantes respectivamente. El tratamiento específico se elige según el estadio del cáncer y la salud general del paciente, con la braquiterapia emergiendo como una opción prometedora en etapas tempranas y avanzadas del cáncer cervicouterino.

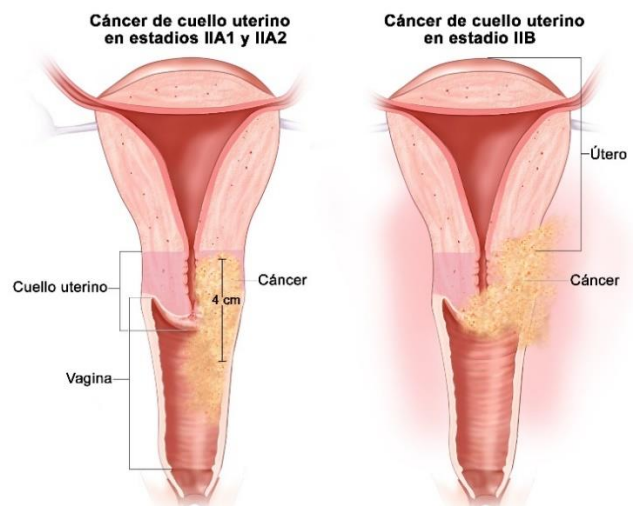

*Fig. 1. cáncer de cervicouterino estadio IIA y IIA2 y IIB*

En presente reporte se analiza el caso clínico de una *paciente femenina de 46 años, con diagnóstico de cáncer cervicouterino en estadio 2B y como paciente de radioterapia externa en 6 campos*. Este caso fue presenciado en el Instituto Oncológico Nacional de la República de Panamá. El ION ha desempeñado un papel fundamental en la lucha contra el cáncer cervicouterino, una de las principales causas de muerte entre las mujeres en el país. Desde su fundación en 1948, el ION ha estado a la vanguardia de la investigación y el tratamiento de diversas formas de cáncer, incluido el cáncer cervicouterino.

## 2. Preparación de la paciente para la recepción de la braquiterapia con Cs-137

Desde el punto de vista como estudiante de Física médica, se observó la estricta seguridad e higiene como protocolo previo a la entrada a la sala de operaciones donde se procede a preparar la paciente para el tratamiento. En primera instancia a la llegada al séptimo piso del ION, se procedió con la colocación del vestuario para el cuarto de cirugía. Fue necesario la implementación de guantes blancos de látex, fundas de zapatos, gorro quirúrgico y el uniforme naranja para hospitales.

Luego de estar con el vestuario correcto, se ingresó a la sala de cirugía donde se ingresó a la paciente entre las 10:20 y 10:30 am. Dentro del personal presente en la sala, estuvo, el radiooncólogo, la anestesióloga, el técnico radiólogo, un técnico instrumentista quirúrgica, una enfermera y el físico médico.

Después de administrar la anestesia a la paciente, se la colocó en la posición de litotomía (*fig. 2*), lo que involucra acostarse boca arriba con las piernas flexionadas y separadas, facilitando así un acceso más adecuado y cómodo al área pélvica y genital para el procedimiento subsiguiente.

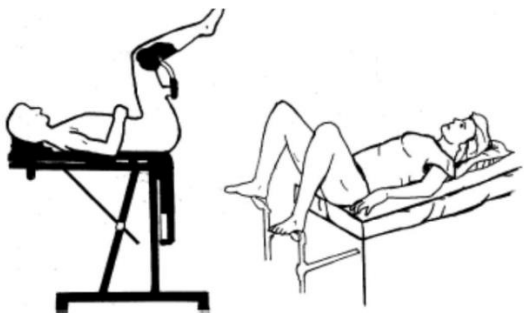

*Fig. 2. Posición de litotomía Forzada.*

El radiooncólogo limpió la vagina meticulosamente con gasas y luego colocó más gasas para proporcionar soporte al tándem vaginal para la inserción de las fuentes radiactivas de cs-137. El instrumento utilizado para abrir la vagina de la paciente y ejecutar el procedimiento se llama especulo vaginal.

Seguidamente, se le colocó a la paciente un marcador rectal diseñado con balines de plomo para ser usado como puntos de referencia en la imagen por radiografía que utilizó el físico médico para la planificación de la dosis.

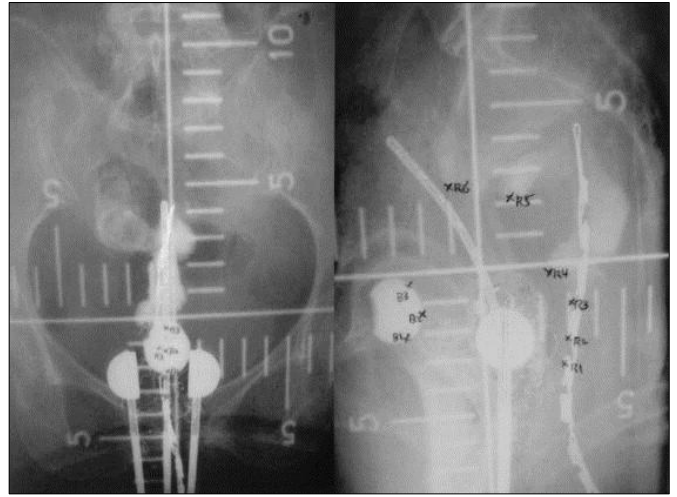

*Fig. 3. tándem vaginal y marcador rectal (International Journal of Radiation Oncology - Biology – Physics, 2003).*

Como se observa en la figura 3, (la cual es una imagen representativa del caso), se puede observar el tándem vaginal que sirve para administrar la dosis y el marcador rectal que sirve de referencia al planificador para que pueda calcular las posiciones correctas en donde se deben aplicar las fuentes de cs-137.

Luego de haber colocado el marcador y el tándem, la paciente se trasladó al cuarto de rayos X para tomar las imágenes que le sirven al planificador para los cálculos.

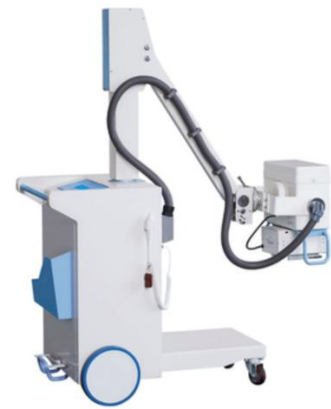

*Fig. 4. Aparato de rayos X.*

En la figura 4 se tiene un ejemplar de un aparato de rayos X similar al usado para la toma de imágenes para la planificación. Durante la toma de las imágenes, los técnicos junto con el físico medico realizaron mediciones de las dimensiones corporales de la paciente como el ancho de la cadera y la altura del área pélvica. Además, realizaron un proceso de ajuste y nivelación del colimador para mejorar la paralaje de los planos de la paciente con los planos de la imagen. Luego de obtenidas las imágenes, el físico medico procedió a calcular las posiciones y a realizar la planificación.

## 3. Planificación de la dosis para el tratamiento

El físico medico se encargó de la planificación de las dosis con un diagrama de la vagina donde marcaba el valor de la fuente y un cuadro con las distancias del plano ortogonal.

El cuadro observado con la información del plano Ortogonal se presenta a continuación:

| Medidas de los planos ortogonales | | |
|---|---|---|
| Parámetro | Plano Axial (cm) | Plano Lateral (cm) |
| Distancia fuente-piel | 80 | 47 |
| Diámetro de la paciente | 19,7 | 38,7 |
| Distancia piel-placa | 2,5 | 15 |
| Distancia fuente-placa | - | - |
| Distancia fuente-marca | 59,32 | 32 |

*Tabla 1. Dimensiones de la paciente para los ajustes de dosimetría.*

En la tabla 1 se puede observar las mediciones hechas por los técnicos y el físico medico en la sala de radiografía. En el proceso de medición los técnicos se encontraron con que las dimensiones de la paciente superaban el limite de la escuadra metálica, calibrada en centímetros. Para solucionarlo, optaron por añadir al extremo de la escuadra una segunda regla calibrada en milímetros.

La precisión en el proceso de medición es sumamente importante debido a que se deben proteger los órganos de riesgo como el recto, la vejiga o cualquier otro tejido que no tenga necesidad de recibir radiación.

Se ajusto el aparato a 10,5 kV para la obtención de las imágenes de rayos X tanto del plano AP, como el plano lateral.

Luego de realizar los cálculos de la dosis se anotó en un diagrama de la vagina las cantidades de dosis y el tipo de fuente, además del tiempo en el que se debía dejar la braquiterapia.

Tanto el radiooncólogo como el físico medico están en constante comunicación para decidir la cantidad de radiación que se le administrara a la paciente en cada uno de los puntos requeridos. Aunque el físico medico tiene la responsabilidad de los cálculos correctos, el radiooncólogo supervisa si se deben realizar pequeños ajustes en la dosis. En la siguiente figura 5 se presenta la distribución de dosis realizada por el físico médico. En el se puede observar como se van a colocar las fuentes y el valor de la cantidad de energía de radiación absorbida por un kilogramo de tejido. Según el diagrama observado en la hoja de prescripción médica, los valores de las fuentes son:

| Dosis (cGray) | 25 | 20 | 15 | 20 | 20 |
|---|---|---|---|---|---|
| Tiempo (horas) | 36 | 26 | 24 | 31 | 32 |

*Tabla 2. Valor de las fuentes para su aplicación a la paciente.*

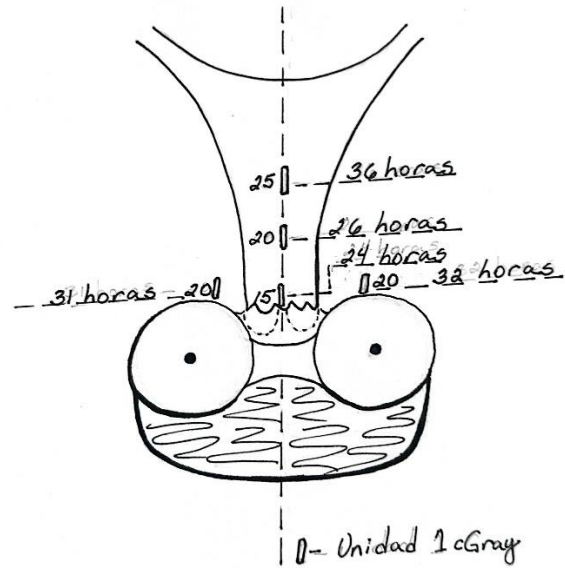

*Fig. 5. Diagrama vaginal de planificación de dosis en braquiterapia de baja tasa con cesio-137 (Dibujado por Ania Edwards, 2023)*

Una ves se tiene la planificación se procede a simular la planificación para ver si existen errores en la cantidad de radiación que la paciente va a recibir, y no ocasionar un accidente.

## 4. Manejo y almacenamiento de las fuentes de Cs-137

Por otro lado, al tener los ajustes necesarios en la dosis para la paciente, existe un técnico en radiología encargado de extraer las fuentes de Cs-137 de una caja de seguridad para fuentes aprobado por la **IAEA** (*Organismo Internacional de Energía Atómica*). Las fuentes se encuentran en pequeñas gavetas que deben estar correctamente blindadas. El técnico debe proceder con todas las normas establecidas de protección radiológica.

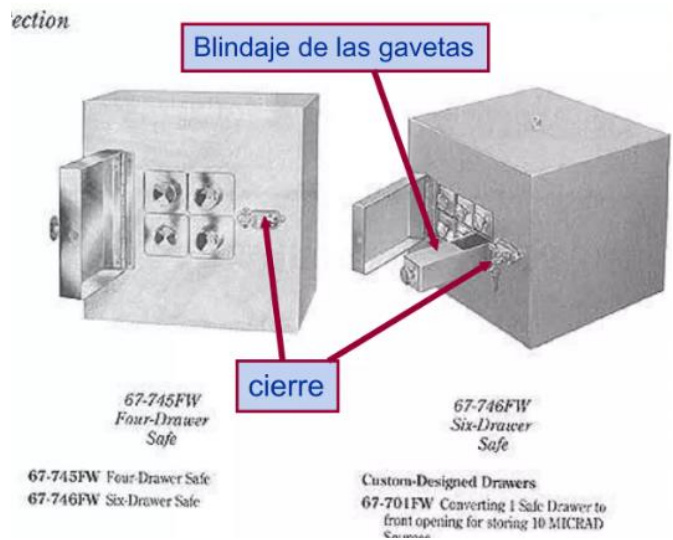

*Fig. 6. Caja de almacenamiento de las fuentes de Cs-137 para braquiterapia de baja tasa.*

Como se puede apreciar en la figura 6, las fuentes están almacenadas debidamente en pequeñas gavetas dentro de la caja de seguridad.

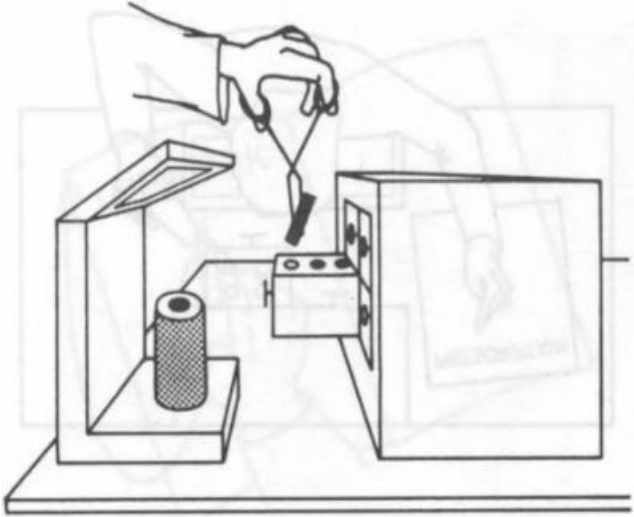

*Fig. 7. Manejo correcto de las fuentes de Cs-137*

La figura 7 es un dibujo representativo sobre el correcto manejo según las normas de seguridad establecidas por la IAEA. El técnico debe pararse detrás de una pared blindada. Luego, debe usar guantes para manejar la caja de almacenamiento de las fuentes y debe tener cuidado de no manipular directamente los cilindros. Debe usar pinzas para sujetar las fuentes, extraerlas de la gaveta y ponerlas en el envase de seguridad. Sus movimientos de trasporte deben ir por la ruta mas corta desde la gaveta hacia el envase de seguridad.

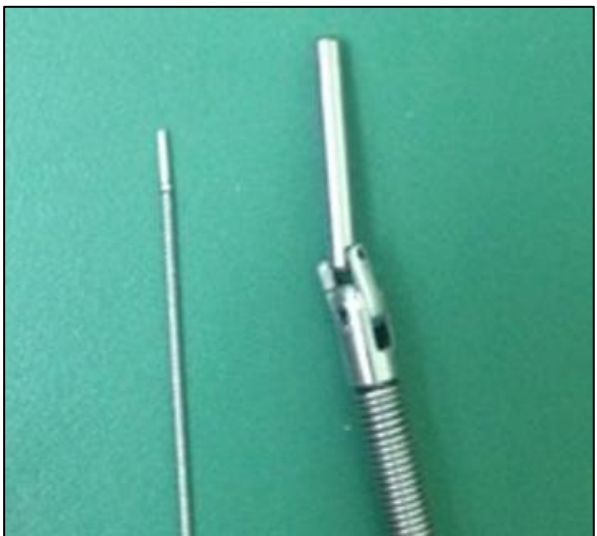

*Fig. 8. Portadores de fuentes radiactivas utilizados en braquiterapia.*

En la figura 8 se muestra una porta fuentes utilizado para llevar las fuentes desde el almacén hasta la sala donde se encuentra la paciente e introducir la fuente en las posiciones correspondiente para el tratamiento. Luego de que el radiooncólogo se encargue de la colocación de las fuentes, en las respectivas posiciones, verificadas por el físico médico, la paciente es trasladada a un cuarto de reposo durante el tiempo necesario para su tratamiento.

## 5. Conclusiones

El proceso de braquiterapia de baja taza con fuentes de Cs-137 observado en el ION, presenta algunas limitaciones en cuanto a la logística para el tratamiento en la paciente. Durante el todo el tratamiento, el uso correcto del vestuario para entrar en la sala de operaciones es de vital importancia, al igual que la desinfección de las manos y el uso de guantes de látex. Se observó el constante cuidado en el manejo de las fuentes por parte del técnico y del físico medico encargado del tratamiento. Se logró un gran aprendizaje en cuanto a la aplicación de los conceptos físicos como los rayos x. Se observó la importancia del trabajo multidisciplinario en un tratamiento de braquiterapia. La colaboración entre varios profesionales de la salud y el ambiente de colaboración hace que el proceso de alta sensibilidad sea prudente y se pueda realizar de manera satisfactoria.

## 6. Referencias

*[1]Barón, L. (s.f.). Posición de Litotomía. Obtenido de uDocz inc:* https://www.udocz.com/apuntes/224106/posicion-de-litotomia

*[2]International Atomic Energy Agency. (27 de noviembre de 2012). Protección radiológica en radioterapia. Obtenido de Exposición Medica: buenas prácticas y protección radiológica en braquiterapia:* https://es.slideshare.net/DanielMellaTreumun/rt11-braqui1aequiposesweb-15380020

*[3]International Journal of Radiation Oncology - Biology - Physics. (s.f.). Obtenido de Unique role of proximal rectal dose in late rectal complications for patients with cervical cancer undergoing high-dose-rate intracavitary brachytherapy:* https://www.sciencedirect.com/science/article/pii/S0360301603007211

*[4]SuQuir C.A. (15 de mayo de 2014). EQUIPO DE RAYOS X PORTÁTIL PLX101D. Obtenido de* https://suquir.com.ve/producto/equipo-de-rayos-x-portatil-plx101d/

*[6]Yáñez, R. (4 de septiembre de 2017). Física de Braquiterapia I. Obtenido de Curso Regional de Capacitación (Actualización en Braquiterapia de Alta Tasa de Dosis):* https://conferences.iaea.org/event/142/contributions/4618/attachments/2954/3536/Physics_of_brachytherapy_I_-_sources_and_delivery_systems.pdf